\RequirePackage[switch, columnwise, running, mathlines, displaymath, mathlines]{lineno}
\RequirePackage{docswitch}
\def\flag{apj}
\setjournal{\flag}

\documentclass[\docopts]{\docclass}

\usepackage{lsstdesc_macros}

\usepackage{graphicx}
\usepackage{url}
\graphicspath{{./}{./figures/}}
\newcommand{\lsst}{\textsc{LSST}}
\newcommand{\plasticc}{\textsc{PLAsTiCC}}
\newcommand{\proclam}{\texttt{proclam}}
\newcommand{\snmachine}{\texttt{snmachine}}
\newcommand{\snphotcc}{\textsc{SNPhotCC}}

\begin{document}

\title{The Photometric LSST Astronomical Time-series Classification Challenge (PLA\MakeLowercase{s}T\MakeLowercase{i}CC): Selection of a performance metric for classification probabilities \\ balancing diverse science goals}

\maketitlepre

\begin{abstract}

  \vspace{0.5cm}

  Classification of transient and variable light curves is an essential step in using astronomical observations to develop an understanding of the underlying physical processes from which they arise.
  However, upcoming deep photometric surveys, including the Large Synoptic Survey Telescope (\textsc{LSST}), will produce a deluge of low signal-to-noise data for which traditional type estimation procedures are inappropriate.
  Probabilistic classification is more appropriate for the data but is incompatible with the traditional metrics used on deterministic classifications.
  Furthermore, large survey collaborations like \textsc{LSST} intend to use the resulting classification probabilities for diverse science objectives, indicating a need for a metric that balances a variety of goals.
  We describe the process used to develop an optimal performance metric for an open classification challenge that seeks to identify probabilistic classifiers that can serve many scientific interests.
  The Photometric \textsc{LSST} Astronomical Time-series Classification Challenge (\textsc{PLAsTiCC}) aims to identify promising techniques for obtaining classification probabilities of transient and variable objects by engaging a broader community beyond astronomy.
  Using mock classification probability submissions emulating realistically complex archetypes of those anticipated of \textsc{PLAsTiCC}, we compare the sensitivity of two metrics of classification probabilities under various weighting schemes, finding that both yield results that are qualitatively consistent with intuitive notions of classification performance.
  We thus choose as a metric for \textsc{PLAsTiCC} a weighted modification of the cross-entropy because it can be meaningfully interpreted in terms of information content.
  Finally, we propose extensions of our methodology to ever more complex challenge goals and suggest some guiding principles for approaching the choice of a metric of probabilistic data products.

  \vspace{1cm}

\end{abstract}

\dockeys{}

\maketitlepost


\clearpage

\section{Introduction}
\label{sec:intro}

The Large Synoptic Survey Telescope (\lsst) will revolutionize time-domain astronomy and the study of transient and variable objects within and beyond the Milky Way.
With its rapid scan strategy, exquisite depth, and multiple optical filters, \lsst\ will deliver millions of light curves, comprised of time-series observations in six electromagnetic wavelength ranges divided into photometric bands in the visible regime.
\lsst's expansive catalog of light curves will enable unprecedented population-level studies of time-varying astrophysical sources, from asteroids to variable stars to active galactic nuclei, deepening our understanding of stellar aging processes, the evolution of the most massive galaxies, and the expansion history of the universe, to name but a few.

Science output from the \lsst\ dataset is, however, contingent on distinguishing classes of astrophysical sources from one another.
Though photometric light curves like those of \lsst\ can be used for classification, costly observations of a high-resolution spectrum have traditionally served as the gold standard for classification.
The volume of objects anticipated of LSST, as well as the potentially low signal-to-noise ratios of the faintest sources, likely exceeds the availability of spectroscopic follow-up resources; the great majority of \lsst's time-varying discoveries will never be spectroscopically confirmed.
As such, there is an acute need for classifiers of photometric light curves that can perform well on datasets that include a wide variety of sources including those that are at the limits of detection.

The Photometric \lsst\ Astronomical Time-series Classification Challenge (\plasticc\footnote{\url{http://plasticcblog.wordpress.com/}, \url{https://www.kaggle.com/c/PLAsTiCC-2018}}) aimed\footnote{\plasticc\ was run as a Kaggle challenge from 17 September 2018 to 17 December 2018.
Though \plasticc\ concluded prior to the final revision of this paper, the study herein was conducted entirely before the commencement of \plasticc, and the draft was submitted to the journal prior to \plasticc's conclusion, hence the use of the present and future tenses throughout this paper.}
to identify and motivate the development of classification techniques that serve astronomical science goals by engaging the broader community outside astronomy.
\plasticc's dataset is comprehensive, including models for well-understood classes, newly observed classes, and classes that have only been proposed to exist, to simulate serendipitous discoveries anticipated of \lsst \citep{the_plasticc_team_photometric_2018, kessler_models_2019}.
Additionally, \plasticc\ joins the ranks of a handful of past astronomy classification challenges including \citep[Mapping Dark Matter\footnote{\url{https://www.kaggle.com/c/mdm}}]{kitching_gravitational_2011}, \citep[Observing Dark Worlds\footnote{\url{https://www.kaggle.com/c/DarkWorlds}}]{harvey_observing_2013}, and \citep[the Galaxy Challenge\footnote{\url{https://www.kaggle.com/c/galaxy-zoo-the-galaxy-challenge}}]{dieleman_rotation-invariant_2015}, all hosted on Kaggle\footnote{\url{https://www.kaggle.com/}}, a platform that hosts data analytics competitions where seasoned professionals and amateurs alike can compete to classify, model, and predict large data sets uploaded by companies or scientific collaborations.
Kaggle attracts a broad userbase, and those without domain knowledge may provide novel approaches to the problem at hand.

Classification in astronomy may proceed through images, as has been done in the contexts of galaxy classification \citep{hoyle_measuring_2016}, supernova classification \citep{cabrera-vives_deep-hits:_2017}, identification of bars in galaxies \citep{abraham_detection_2018}, weak lensing estimation\footnote{\url{http://great3challenge.info/}}\citep{mandelbaum_third_2014}, separation of Near Earth Asteroids from artifacts in images \citep{morii_machine-learning_2016}, as well as  time-domain classification \citep{morii_machine-learning_2016, mahabal_deep-learnt_2017, zevin_gravity_2017}, and even noise classification \citep{zevin_gravity_2017, george_classification_2018}.
Classification may also proceed from time-series or spectroscopic data rather than images, as in \citet{newling_statistical_2011, richards_construction_2012, ishida_kernel_2013, richards_bayesian_2015, armstrong_k2_2016, lochner_photometric_2016, moller_photometric_2016}.
Automated classification \citep{mahabal_automated_2008, djorgovski_towards_2011, bloom_automating_2012, djorgovski_flashes_2012, narayan_machine_2018} is becoming increasingly important in time-domain astronomy due to its potential for speed relative to visual inspection by an expert; the sooner one can make follow-up observations of an interesting object, the more one can learn about its underlying physical processes and nature.

Classification is intrinsically \textit{probabilistic} in that the goal is to constrain the class \textit{conditioned} on limited data, thereby defining a \textit{posterior probability density}, or \textit{classification posterior} for short, over all classes for each classified light curve.
Probabilities of classification that are reduced to an estimated class label (say, by rounding a probability $0 \leq p \leq 1$ up or down) without a notion of confidence become \textit{deterministic} classifications.
Such a reduction of a probability density to a deterministic label discards information, the impact of which depends on how the classification results are subsequently used.

Probabilistic classifications could inform decisionmaking regarding allocation of limited spectroscopic follow-up resources.
To reduce wasting spectroscopic resources dedicated to a common class whose science use requires spectra, one might only attempt follow-up observations of the objects with the highest classification probabilities.
Spectroscopic follow-up of a rare class, on the other hand, may be useful enough that an object with even a moderate probability of being of a very rare class could be worth the risk.

Perhaps more significantly, classification probabilities may be propagated through a hierarchical inference of population-level parameters, enabling scientific investigations to proceed even when spectra are unavailable.
The efficacy of this application of classification probabilities in the context of supernova cosmology is an active field of research \citep[Malz, Peters, and Hlo\v{z}ek in prep]{rubin_unity:_2015, roberts_zbeams:_2017, jones_measuring_2018}.
Thus the impact of a photometry-only survey like \lsst\ can be greatly enhanced by probabilistic classifications.

In light of the aforementioned benefits of classification probabilities, \plasticc\ will thus accept classifiers producing classification posteriors.\footnote{Classifiers that only provide deterministic or binary classifications (including some of the most prevalent classifiers in the field of time-domain astronomy) will have to convert their results to probability vectors to compete in \plasticc.}
However, probabilistic classifications are incompatible with the \textit{metrics}, any quantification of the performance of a classifier, of deterministic label assignments used in previous classification challenges \citep{kessler_supernova_2010, kessler_results_2010} and efforts to develop supernova classifiers \citep{narayan_machine_2018}.
Accuracy, purity, completeness, and contamination are examples of metrics of deterministic classification estimates that are commonly used in astronomical applications.

Many deterministic classification metrics can be modified for evaluation on classification posteriors \citep{gieseke_detecting_2010, lochner_photometric_2016, moller_photometric_2016, hon_deep_2017, hon_detecting_2018}, but only by reducing class probabilities to deterministic labels via evaluation at different cutoffs, the choice of which may ultimately affect the value of the metric and thus assessment of the classifier.
Furthermore, many such metrics are restricted to binary classifications (``yes'' or ``no'') and thus do not meet the diverse needs of \plasticc.

If the data are simulated using a fully self-consistent forward model, a metric of the accuracy of classification posteriors relative to the true, underlying probabilities would be straightforward.
However, such a simulation procedure would require beginning with a fully populated probability space over all classes and all possible light curves, which is an insurmountable challenge.
Therefore, attention must be directed toward defining the criterion for identifying a winning classifier.
In the context of astronomy, concerns about the choice of metric for probabilistic classifications have been investigated \citep{kim_stargalaxy_2017, florios_forecasting_2018}, though most studies focus on the standard metrics of purity and completeness.
Even within that subset, metric consistency over a range of classifiers and between different analyses is not always ensured \citep{bethapudi_separation_2018}, indicating a need for further study.

This work explores the problem of how to choose a metric of probabilistic classifications with intended application to many science applications.
The \plasticc\ metric must respect the information content of probabilistic classifications without reduction to point estimates of class;
it must be well-defined for non-binary classes, going beyond a positive/negative dichotomy inherent to some traditional metrics.
The winning classifier should not favor one science application above all others, necessitating robustness against significant class imbalance, both between and within the training set and test set, as well as other concerning systematics.
Finally, in order for the metric to satisfy the challenge requirements, the metric must return a single, scalar value.

We perform a systematic exploration of the sensitivity of metrics of probabilistic classification to anticipated classifier failure modes using the PRObabilistic CLAssification Metric (\proclam) code \citep{malz_proclam_2018}, which is publicly available on GitHub\footnote{\url{https://github.com/aimalz/proclam}}.
The mock classification submissions that we use for this study are described in Section~\ref{sec:data}.
The metrics we consider are presented in Section~\ref{sec:methods}.
The behavior of the metrics as a function of mock classification results is presented in Section~\ref{sec:results}.
We discuss extensions of this exploratory framework to more complex challenge goals in Section~\ref{sec:discussion}.

\section{Data}
\label{sec:data}

We explore the behavior of metrics on mock classification probabilities with isolated strengths and weaknesses as well as realistic mock classification probabilities from a publicly available light curve catalog.
Throughout this paper, \textit{data} always refers to mock classification submissions to \plasticc, not the \plasticc\ light curves; no light curves were simulated, viewed, or classified in the preparation of this paper.

Our data is in the form of catalogs of $N$ posterior probability vectors $p(m \mid d_{n}, D, \mathcal{C})$ over $M$ classes with labels $m$ conditioned on each observed light curve $d_{n}$, the training set $D$, and some parameters $\mathcal{C}$ concerning the behavior of the classifier.
We motivate $\mathcal{C}$ here before deferring its detailed explanation to later in Section~\ref{sec:mockdata}.

If a mock classifier produced $p(m \mid d_{n})$, it would take solely the light curve and produce a posterior over classes.
Since such a situation involves no information besides the light curve $d_{n}$, every classifier would produce identical classification submissions $\bar{p}(m \mid d_{n})$.
Including the training set $D$ would not remedy the problem, as every classifier for \plasticc\ has access to the same training set and so would still have no way to produce different classification submissions $p(m \mid d_{n}, D)$.
Thus there must be some other parameters $\mathcal{C}$ that are specific to each classifier and contribute to the mock classification posteriors it produces.\footnote{It should be noted that classification submissions may not be derived in this way, i.e. the parameters $\mathcal{C}$ may not be explicitly known or may indicate a procedure that does not produce posteriors but, rather, scores of some kind.  However, we assume for these purposes that classifiers produce the classification posteriors \plasticc\ seeks.}
We describe below the way in which mock data is synthesized and return to the classifier parameters $\mathcal{C}$ later.

\begin{figure}
	\begin{center}
    \includegraphics[width=0.49\textwidth]{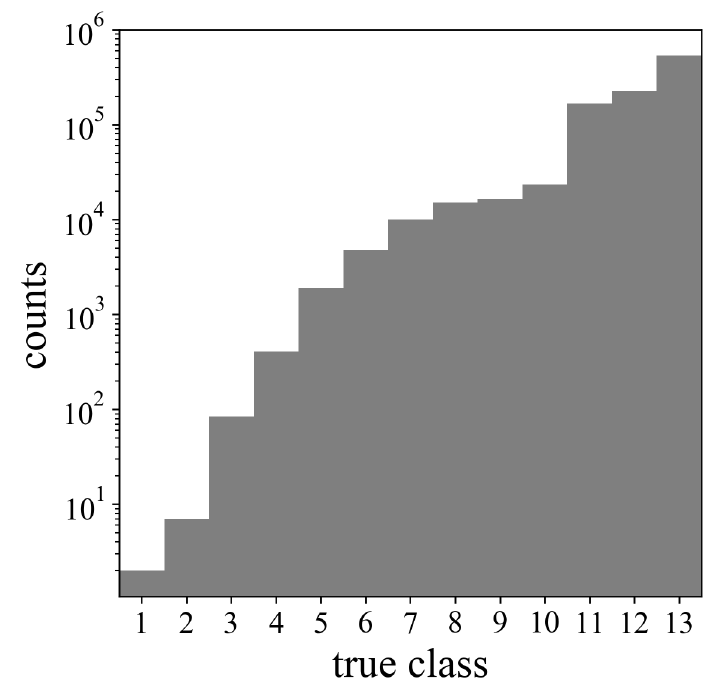}
		\caption{The number of members of each of thirteen mock classes considered in this work.
		Class populations were simulated by drawing the number of members of a given class from a logarithmic distribution to emulate the extreme class imbalances typical of astronomical samples.}
		\label{fig:classdist}
	\end{center}
\end{figure}

As is anticipated of the real \lsst\ dataset, we use class populations that are logarithmically distributed such that they span many orders of magnitude.
We then take $M$ draws $u_{m} \sim \mathcal{U}(0, 1)$ from the standard continuous uniform distribution.
These draws $\{u_{m}\}$ are used to establish a discrete probability distribution $p(m) = b^{u_{m}}\ / \sum_{m} b^{u_{m}}$ such that $\sum_{m=1}^{M} p(m) = 1$.
From $p(m)$ we draw $N = 10^{b}$ instances $\{m'_{n}\}$ of a true class $m'$ for each light curve $n$ in the catalog.

The true class membership distribution of our tests with $M = 13$ and $b = 6$ is shown in Figure~\ref{fig:classdist}.
Though the class labels for \plasticc\ are expected to be randomized, we artificially order our mock class labels by their prevalence for ease of visual interpretation.
Once the true classes have been set, mock classification probabilities for each class are derived using the procedure described in Section~\ref{sec:mockdata}.

\subsection{Mock classification schemes}
\label{sec:mockdata}

In order to observe metric performance on different classification schemes, we simulate some archetypical mock classifiers, devised to produce generic responses to a classification challenge, without any interaction with actual challenge data, nor any other light curves.
We use these mock classifiers to investigate how the performance under each metric changes in the presence of certain types of failure modes, or \textit{systematics}.
A robust metric should not reward classification schemes that display these systematic effects.

The archetypical systematics can be seen as modifications to the confusion matrix, a measure of deterministic classification \citep{bloom_automating_2012}.
The confusion matrix is an $M \times M$ table of observed counts (or, if normalized, rates) of pairs of estimated class labels $\hat{m}$ (columns) and true classes $m'$ (rows) computed after a deterministic classification has been performed on some data set with $N$ objects.

Under a binary deterministic classification between positive and negative possibilities, the confusion matrix contains the numbers of true positives $\mathrm{TP}$, false positives $\mathrm{FP}$ (Type 1 error), true negatives $\mathrm{TN}$, and false negatives $\mathrm{FN}$ (Type 2 error), which can be turned into rates relative to the true numbers of positive and negative instances.
These rates may serve as building blocks for more sophisticated metrics of multi-class deterministic classifiers addressed in Section~\ref{sec:methods}.
Though probabilistic classifications are not compatible with the confusion matrix, regardless of normalization, we design tests around proposed normalized confusion matrices exhibiting various systematics that we anticipate being problematic for \lsst.

Under a deterministic classification scheme with a normalized confusion matrix with elements $p(\hat{m}, m')$, an object with true class $m'$ would have an assigned class $\hat{m}$ drawn from $p(\hat{m} \mid m') = p(\hat{m}, m') / p(m')$, via Bayes' Rule.
We note that the elements of the confusion matrix have values of $N p(\hat{m}, m')$ and that $p(m') = N_{m'} / N$, where $N_{m'}$ is the number of true members of class $m'$, must be known in order to produce a confusion matrix.
We refer to the matrix $\mathbb{C}$ composed of $p(\hat{m} \mid m')$ as the \textit{conditional probability matrix} (CPM), and we use it to derive mock classification posteriors.

Assuming the light curves contain information about the true class (an assumption that underlies classification as a whole), we can use the appropriate row $\mathbb{C}_{m'_{n}} = p(\hat{m} \mid m', \mathcal{C})$ of the CPM $\mathbb{C}$ as a proxy for $p(m \mid d_{n}, D, \mathcal{C})$, without directly classifying light curves themselves.\footnote{This assumption is key to the generality of this work, which was conducted without any knowledge of the \plasticc\ dataset simulation procedure.}
To emulate the effect of natural variation of information content in different light curves (e.g. a noisy lightcurve has less information to recover than one with a higher signal-to-noise ratio) using the above, we generate a posterior probability vector $\vec{p}(m \mid m', \mathbb{C})$ by taking a Dirichlet-distributed draw
\begin{eqnarray}
  \label{eq:cmtoprob}
  \vec{p}(m \mid d_{n}, D, \mathcal{C}) &\sim& \mathrm{Dir}[\mathbb{C}_{m'_{n}} \delta]
\end{eqnarray}
about $\mathbb{C}_{m'_{n}}$, with a small nonnegative perturbation factor $\delta = 0.01$.
In this way, the posterior probability vector has an expected value equal to the appropriate row in the CPM, with a variance set by $\delta$.
We impose one restriction in addition to the normalization factor of Equation~\ref{eq:cmtoprob}, namely that all elements of $p(m \mid d_{n}, D, \mathcal{C})$ exceed $10^{-8}$, to ensure numerical stability in light of the limitations of floating point precision.

\begin{figure*}
	\begin{center}
    \includegraphics[width=0.8\textwidth]{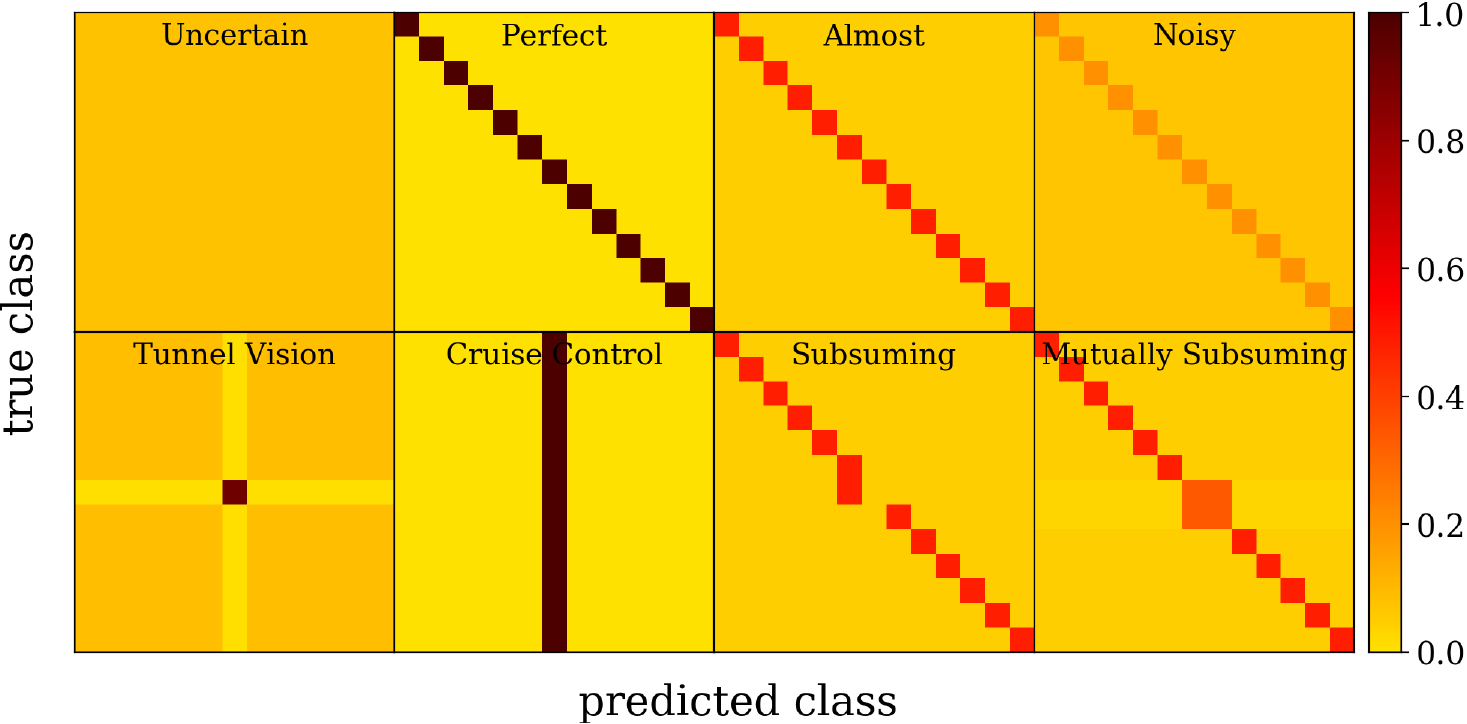}
		\caption{Conditional probability matrices (CPMs) for eight mock classifiers.
		Top row:
		the uncertain classifier's uniform CPM;
		the perfect classifier's identity CPM;
		the almost perfect classifier's CPM, a linear combination of one part uniform and four parts identity;
		the noisy classifier's CPM, a linear combination of one part uniform and two parts identity.
		Bottom row:
		the tunnel vision classifier's CPM is uniform except at the row and column corresponding to one class, where it takes the values of the identity matrix;
		the cruise control classifier's CPM, which has the every row equal to a particular row of the identity;
		the subsuming classifier's CPM, which has two or more rows equal to one another;
		the mutually subsuming classifier's CPM, a symmetric case of the subsuming classifier.
		The top row shows CPMs that serve as unbiased control cases.
		The CPMs of the bottom row represent concerning systematics that we would like to ensure are not rewarded by the \plasticc\ metric.
		}
		\label{fig:mock_cm}
	\end{center}
\end{figure*}

We consider eight mock classifiers, each characterized by a single systematic affecting their CPM.
Figure~\ref{fig:mock_cm} shows the CPMs corresponding to each systematic considered, discussed in detail below.

For each of our archetypical mock classifiers, we address:
\begin{enumerate}
  \item What characteristic behavior defines this classifier?
  \item Under what conditions does this behavior arise in real classifications?
  \item What are our expectations of and desires for response of the metric to this archetypical classifier?
\end{enumerate}

\begin{figure}
	\begin{center}
		\includegraphics[width=0.49\textwidth]{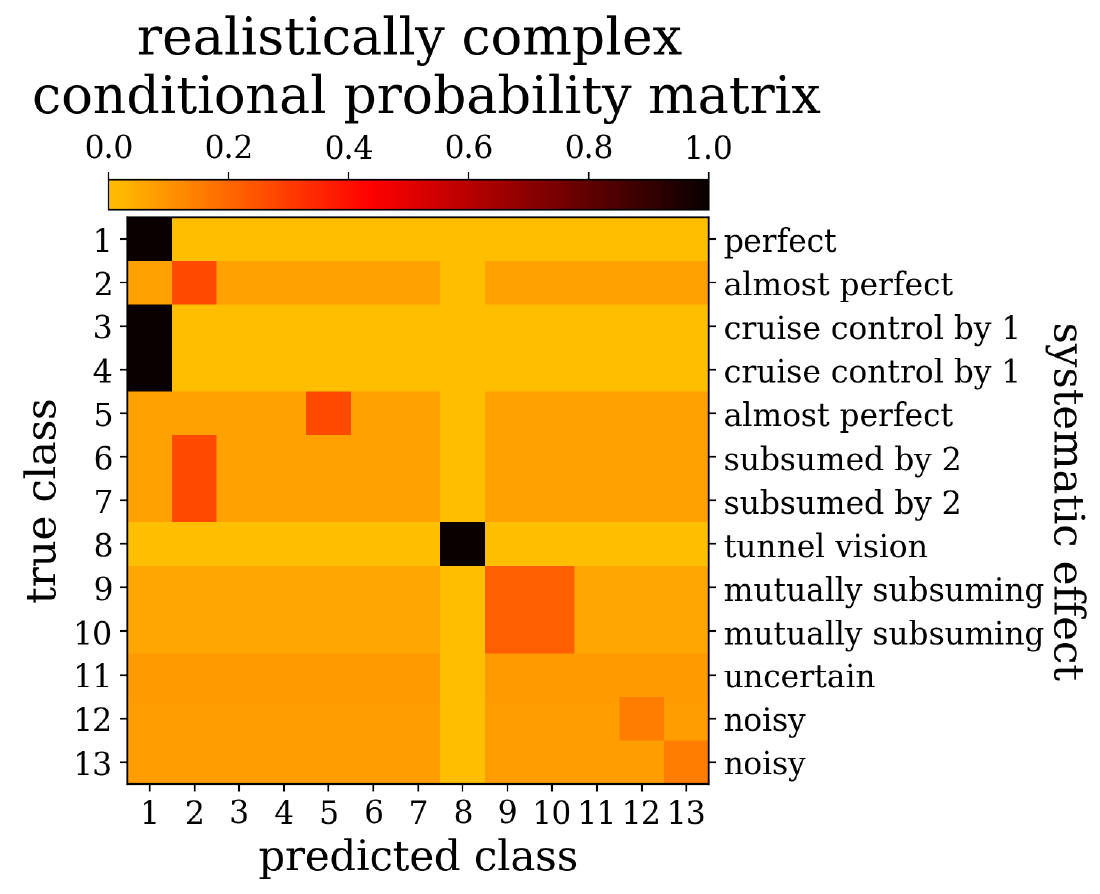}\\
    \includegraphics[width=0.49\textwidth]{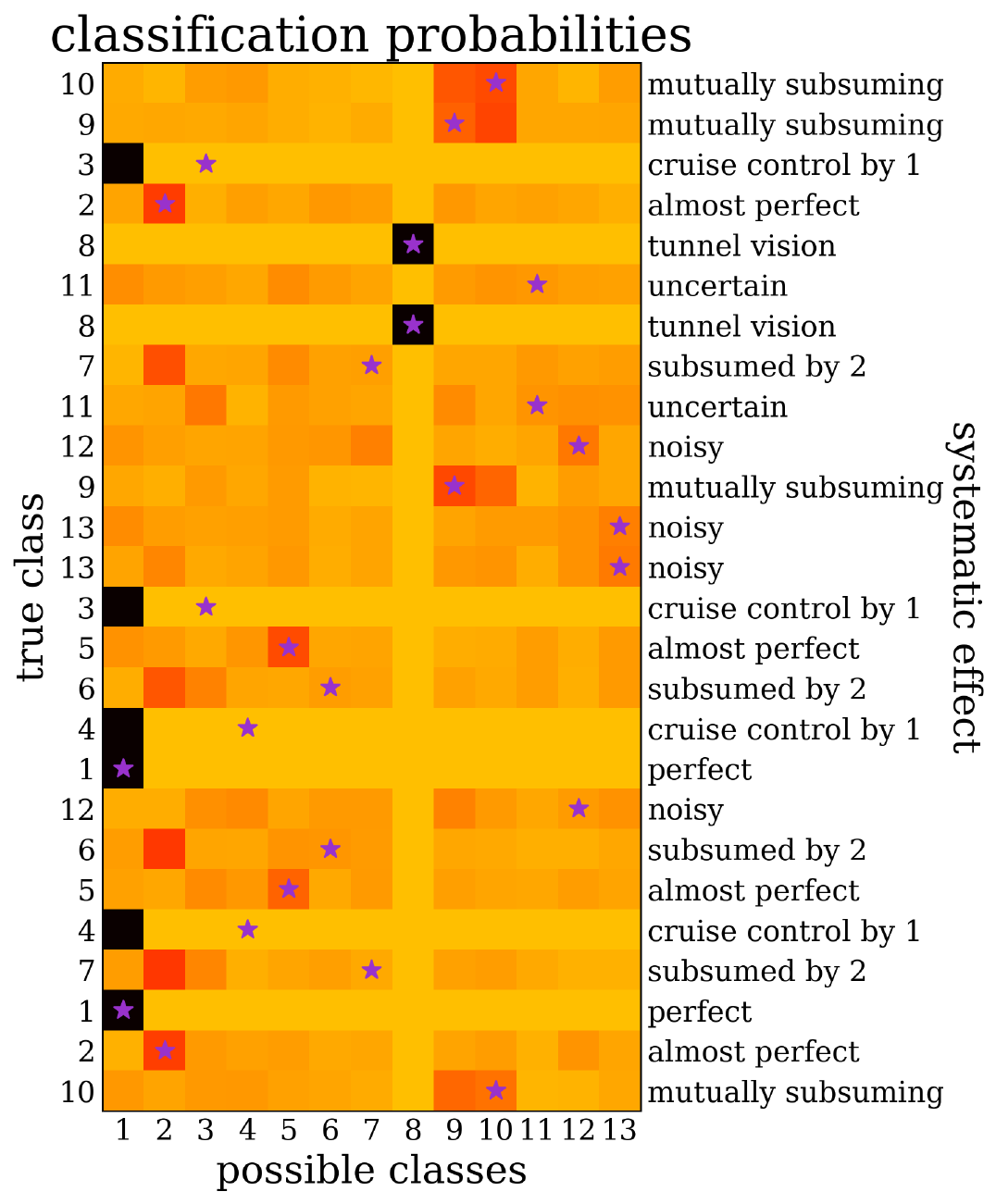}
		\caption{A realistically complex conditional probability matrix (CPM) and classification posteriors drawn from it.
		Top: An example of a realistically complex conditional probability matrix, constructed by selecting a systematic for each individual class.
		This illustrates (for example), how a classifier may exhibit multiple systematics from Figure~\ref{fig:mock_cm} for each true class.
		Bottom: Example classification probabilities, drawn from the above CPM, with their true class indicated by a star and the systematic, characterized by its row in the CPM, affecting that true class described on the right.
		The Dirichlet process emulates the variation in classification posteriors due to differences between light curves within a given class, leading to different classification posteriors even among rows sharing a true class.
		}
		\label{fig:mock_probs}
	\end{center}
\end{figure}

An actual classifier is expected to be more complex than the simplified cases of Figure~\ref{fig:mock_cm}, with different systematic behavior for each class.
An example of a combined CPM across different classes and systematics is given in the top panel of Figure~\ref{fig:mock_probs}.
The rows of this CPM correspond to rows of the archetypical classifiers of Figure~\ref{fig:mock_cm}.
To demonstrate the procedure by which mock classification posteriors are generated from rows of the CPM, we provide 26 examples of draws of the posterior CPM in the bottom panel of Figure~\ref{fig:mock_probs}.
Given a set of true class identities, the mock classification posteriors of the bottom panel are Dirichlet draws from the corresponding row of the CPM of the top panel.

\subsubsection{Uncertain classification}
\label{sec:uncertaindata}

A CPM $\mathbb{U}$ with uniform probabilities for all classes, as shown in the leftmost top panel of Figure~\ref{fig:mock_cm}, would correspond to uniform random guesses for deterministic classification, but in accordance with Equation~\ref{eq:cmtoprob}, the classification posteriors are perturbations away from a uniform distribution across all classes.
The peak values of one such classification posterior would correspond to random classification drawn from a uniform distribution, with $p(m' \mid d_{n}, D, \mathcal{C}_{\mathbb{U}}) \approx M^{-1}$.
We can consider the \textit{uncertain} classifier as an experimental control for the least effective possible classification scheme, bearing in mind that if classifications were anticorrelated with true classes, the experimenter could simply reassign the classification labels to improve performance under any metric.

\subsubsection{Accurate classification}
\label{sec:accuratedata}

The \textit{perfect} classifier has a diagonal CPM $\mathbb{I}$ (left-center top panel of Figure~\ref{fig:mock_cm}), which would correspond to deterministic classifications that are always correct.
In terms of probabilistic classifications, a perfect result would be a classification posterior with 1 for the true class and 0 for all other classes.
In accordance with the classification posterior synthesis scheme of Equation~\ref{eq:cmtoprob}, the class with maximum probability is almost always still the true class, and indeed with $N \sim 10^{6}$ and $\delta = 0.01$, this is always true.
This case is also a control, in that \plasticc\ would not be necessary if we believed the perfect classifier were potentially achievable.

In addition to a perfect classifier, we test linear combinations $\mathbb{C} = (s + 1)^{-1} \left(s\mathbb{I} + \mathbb{U}\right)$ of the perfect and uncertain CPMs where the contribution of the perfect classifier is greater than that of the uncertain classifier by a factor of $s > 0$.
Deterministic classifications drawn from such a CPM would be correct $s$ times as often as they take any one wrong label, and the incorrect labels would be uncorrelated across classes.
The classification posteriors drawn from such CPMs would have some probability at classes other than the true class, but almost all would still have their peak value at their true class.
We consider the case of the \textit{almost perfect} classifier with $s=4$ (right-center top panel of Figure~\ref{fig:mock_cm}) and the \textit{noisy} classifier with $s=2$ (rightmost top panel of Figure~\ref{fig:mock_cm}).

A classifier with different accuracy for each class may be considered a systematic in its own right.
An extreme example of such a classifier is one with perfect classification performance on one class and uncertain classification on all others.
This classifier's CPM would be uniform except for one row, which would take a value of unity on the diagonal and zero elsewhere; if the classifier were also resilient against Type 1 errors, the CPM would also take zeros along the column in question, aside from the value of unity on the diagonal.
For a single science application, this type of classifier is desirable, but the goal of \plasticc\ is to serve the needs of those who study a wide variety of classes for different purposes.
Hence, from the perspective of \plasticc, we seek a metric that disfavors the \textit{tunnel vision} classifier (leftmost bottom panel of Figure~\ref{fig:mock_cm}).

\subsubsection{Inaccurate classification}
\label{sec:inaccuratedata}

If a deterministic classifier is systematically inaccurate, its CPM has significant off-diagonal contributions.
We model inaccurate probabilistic classifications of class $m'$ by using the row of the CPM corresponding to class $\tilde{m}$ as the basis for the perturbed probability vector $p(m \mid m') = p(m \mid \tilde{m})$.
Class $m'$ is said to be \textit{subsumed} by class $\tilde{m}$ by a classifier that absorbs class $m'$ into class $\tilde{m}$ (right-central bottom panel of Figure~\ref{fig:mock_cm}).
The subsuming classifier may be asymmetric, or the classes may be mutually subsumed (rightmost bottom panel of Figure~\ref{fig:mock_cm}) if one already has significant off-diagonal probability, as is true for the uncertain classifier.

Subsuming is not always the mark of a poor classifier and may be insurmountable by more sophisticated classification techniques.
Real classification posteriors $p(m \mid d_{n}, D, \mathcal{C})$ are conditioned on light curves, training data, and assumptions necessary for the classification algorithm, and there may simply not be enough information in a light curve and/or training set to distinguish between classes.

For example, based on only the first few light curve points, it is sometimes impossible to separate cataclysmic variables (stars that are not destroyed and can brighten and fade many times) from supernovae, which are stars that are completely destroyed in their explosions.
Even with observations over extended periods, it can still be impossible to distinguish cataclysmic variables from active galactic nuclei that result from activity near a galaxy's central black hole.
Similarly, tidal disruption events that occur when stars are destroyed by proximity to the central black hole of a galaxy can look much like supernovae that simply happen to be near a galaxy's center.
When the prior information of the location of the source is more informative than its sparse, noisy, irregularly sampled, or short light curve, it may present a challenge no classifier can overcome, a fundamental limit on available information about the object.

Distinguishing between subclasses of a single phenomenon is subject to limits not only on the light curves of the unknown targets but also by the availability of adequate training sets.
It is nonetheless essential to identify subclasses when they have wholly different science applications.
As an example, supernovae (SN) Ia and Ibc are notorious for being difficult to distinguish.
In fact, it is more common for SN Ibc to be misclassified as SN Ia than the other way around.
This asymmetry is due to systematic underrepresentation of SN Ibc in available training sets.
However, SN Ibc contaminants in the traditional cosmology analysis done with SN Ia can bias estimates of the cosmological parameters, so the distinction is critical.

Class imbalance is a ubiquitous problem in astronomy that can severely exacerbate this form of inaccuracy, as the relative rates of various astrophysical events and objects differ by orders of magnitude from one another.
For example, RRc and RRd Lyrae stars are challenging to separate despite having different pulsation modes, and RRd stars, due to their rarity, are typically subsumed by RRc labels.

An extreme case of inaccurate classification is to classify all objects as the most common class (in the training or test set), which is of particular concern to \plasticc\ given non-representative class balance of the training set.
Such a \textit{cruise control} classifier (left-center bottom panel of Figure~\ref{fig:mock_cm}) counters \plasticc's goal of identifying objects belonging to extremely rare classes.
We would like the \plasticc\ metric to reward a classifier that successfully avoids this kind of error.

\subsection{Realistic classifications}
\label{sec:realdata}

\begin{figure*}
	\begin{center}
    \includegraphics[width=\textwidth]{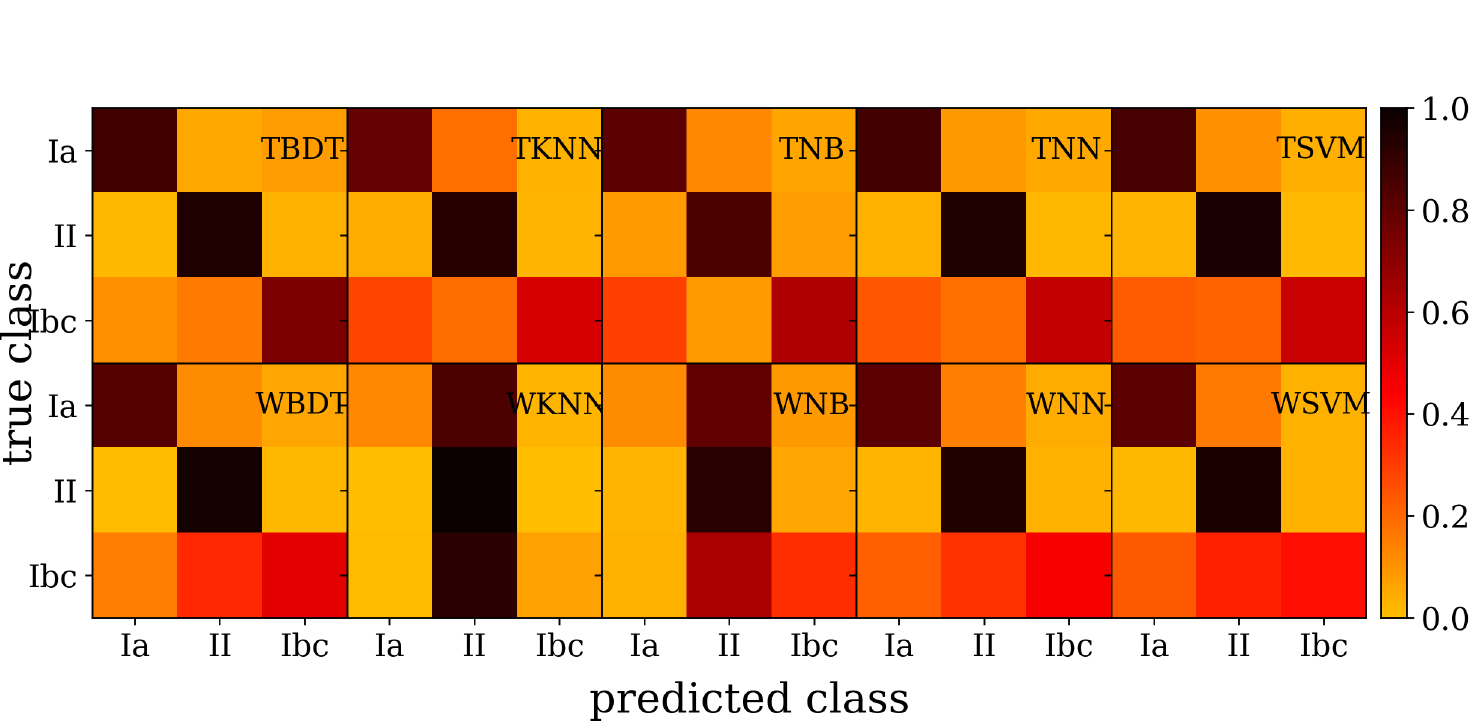}
		\caption{Conditional probability matrices (CPMs) of the \citet{lochner_photometric_2016} methods applied to the second post-challenge release of the \snphotcc\ dataset.
		Columns: the five machine learning methods of Boosted Decision Tree (BDT), K-Nearest Neighbors (KNN), Naive Bayes (NB), Neural Network (NN), and Support Vector Machine (SVM).
    Top row: five machine learning methods applied to template decompositions as features.
    Bottom row: the same five machine learning methods applied to wavelet decompositions as features.
		These CPMs derived from the dataset of a precursor light curve classification challenge by modern methods exhibit some of the systematics identified in Section~\ref{sec:mockdata} and Figure~\ref{fig:mock_cm}, particularly cruise control (WKNN, WNB), noisy (class Ibc in all but TBDT and WKNN), and perfect (class II in all).
		It is worth noting that \citet{lochner_photometric_2016} applies their classification to a representative sub-sample of the \snphotcc\ data selected once the challenge was complete, circumventing some of the issues of non-representativity present in the original submissions to the \snphotcc.}
		\label{fig:snphotcc_cm}
	\end{center}
\end{figure*}

In order to understand the performance of classifiers on simulated datasets approximating reality, we calculate the values of our metric candidates on representative classifiers of a precursor light curve classification challenge.
The Supernova Photometric Classification Challenge (\snphotcc) \citep{kessler_supernova_2010} focused on deterministically classifying a heterogenous population of supernovae into subclasses of SN Ia, SN II, and SN Ibc.

The \snphotcc\ attracted diverse classification approaches, encompassing $\chi^{2}$ fits of the supernova light curves to publicly available templates \citep{nugent_kcorrections_2002}, empirical models \citep{conley_sifto:_2008}, as well as alternatives to curve-fitting such as outlier identification on the training set Hubble diagram, dimensionality reduction, and clustering.
Machine learning was also employed, using features such as the light-curve slopes to produce a predictive model for the training data.

Since the conclusion of the \snphotcc, the light curves became a testbed for a suite of machine learning classifiers.
We consider a collection of probabilistic classification methods, as presented in \citet{lochner_photometric_2016}, whose CPMs\footnote{The classifiers of \citet{lochner_photometric_2016} are indeed probabilistic but are reduced to confusion matrices via deterministic labels (by assigning a label of the class achieving the highest probability) for this visualization and the science-motivated metric of Section~\ref{sec:deterministic}.
In all other instances, the classification posteriors are used directly.} are shown in Figure~\ref{fig:snphotcc_cm}.

The set of classification algorithms includes template-based classification procedures, denoted as T, (\citet{sako_photometric_2011}, top row) and a wavelet decomposition, denoted as W, of the light curves to construct the features over which to classify (\citet{newling_statistical_2011}, bottom row), each paired with Boosted Decision Tree (BDT), K-Nearest Neighbors (KNN), Naive Bayes (NB), Neural Network (NN), and Support Vector Machine (SVM) machine learning algorithms (columns).
While the complexity of entries to the \snphotcc\ was greater than this subset, we use these examples to establish the behavior of our metrics on realistic classification submissions.

We draw attention to the marked presence of the systematics introduced in Section~\ref{sec:mockdata} in the CPMs of Figure~\ref{fig:snphotcc_cm}.
Note that the WNN and WNB methods both suffer from the cruise control systematic on SN II, which were the most prevalent in the \snphotcc\ dataset.
Nearly all the other CPMs exhibit classifications that are almost perfect for SN Ia, perfect for SN II, and noisy for SN Ibc.
A likely cause for this effect is that SN Ibc are poorly represented in training and template sets.

\section{Methods}
\label{sec:methods}

To optimally discriminate between classification techniques, there must be a performance metric, a single scalar value quantifying how appropriate a classifier is for the task at hand.
Choosing a metric for \plasticc\ therefore is logically entwined with the challenge goals.

In Section~\ref{sec:deterministic}, we review a familiar binary, deterministic metric of light curve classification in astronomy.
In Section~\ref{sec:probabilistic}, we introduce metrics appropriate for multi-class probabilistic classification.
We take weighted averages of the per-object metrics with per-class weights described in Section~\ref{sec:weights}.

\subsection{Science-motivated deterministic metric}
\label{sec:deterministic}

We begin with a presentation of a classification metric that has been used in the evaluation of astronomical light curve classifiers in the recent past.
The metric we highlight makes use of the notions of true positive, false positive, and false negative counts from binary deterministic classification.
We briefly define the \textit{efficiency} $\epsilon \equiv \mathrm{TP} / (\mathrm{TP} + \mathrm{FN})$ and \textit{purity} $\pi \equiv \mathrm{TP} / (\mathrm{TP} + \mathrm{FP})$.

The goal of the \snphotcc\ was to identify one particular type of astrophysical source, SN Ia, for a single scientific application, cosmology.
As the \snphotcc\ was only concerned with SN Ia cosmology, it was effectively binary, in that the metric did not distinguish between non-Ia classes.
Since the only SN Ia that would be considered for a cosmology analysis at the time were those with spectroscopic redshifts, the classification was not only binary but also deterministic.
The \snphotcc\ metric $\mathrm{FoM} \equiv \epsilon \cdot \tilde{\pi}$ is the product of the efficiency of SN Ia classification and a modification $\tilde{\pi} \equiv \mathrm{TP} / (\mathrm{TP} + r \mathrm{FP})$ of the purity in terms of a penalty factor $r$.
The inclusion of this second term was motivated by the potential impact on cosmological parameter constraints due to contamination of the SN Ia sample by non-Ia classes.
The pseudo-purity can be interpreted as the traditional purity when $r = 1$ as it is related to the size of the spectroscopic sample; for the \snphotcc, $r=3$ was used.

\subsection{Probabilistic metrics}
\label{sec:probabilistic}

In contrast to \snphotcc's sole goal of optimal deterministic classification of a single class, \plasticc\ seeks to identify classifiers that produce multi-class classification posteriors.
We consider two metrics of classification probabilities that avoid reducing probabilities to deterministic labels.

Our probabilistic metrics are composed of quantities defined for each possible class $m$ among $M$ potential classes available to light curve $n$, which is a true member of the set $\mathbb{S}_{m'}$ of astrophysical sources of class $m'$.
The metric value $Q_{n} = \sum_{m=1}^{M} Q_{n, m}$ for a single light curve $n$ is a sum of the per-class per-light curve metric values $Q_{n, m}$.
The metric value $Q_{m'} = \sum_{n \in \mathbb{S}_{m'}} Q_{n}$ for an entire class $m'$ is the sum of the per-light curve metrics.
Section~\ref{sec:weights} discusses how the global metrics are derived from the per-class metrics $Q_{m'}$.

As part of the derivation of the per-class per-light curve metrics, we also define the indicator variable
\begin{eqnarray}
  \label{eq:indicator}
  \tau_{n, m} &\equiv& \begin{cases}
  0 & m' \neq m\\
  1 & m' = m
  \end{cases}
\end{eqnarray}
that indicates if an object has been correctly classified as its true type.

\subsubsection{Log-loss}
\label{sec:logloss}

The log-loss is a quantity borrowed from information theory and is related to a notion of \textit{entropy} $H_{n} = - \sum_{m=1}^{M} p(m \mid d_{n}) \ln[p(m \mid d_{n})]$, a measure of the space of possible states a system can have, which is in this case the class of which a light curve can be a member.
A classification posterior $p(m \mid d_{n})$ has minimal entropy if it takes a value of $1$ at some class and values of $0$ at all others, i.e. if it can trivially be reduced to a deterministic classification, because this is the scenario in which there is only one possible state, that the light curve has a true class $m$.
This definition of entropy, however, is a property of the probability $p(m \mid d_{n})$ and has no relation with any concept of the true class of the light curve $m'$.

To reconcile the classification posterior with the true class known by those running a challenge, we define the cross-entropy
\begin{eqnarray}
  \label{eq:logloss}
  L_{n} \equiv Q^{L}_{n} &=& -\sum_{m=1}^{M} \tau_{n, m} \ln[p(m \mid d_{n})],
\end{eqnarray}
which can be interpreted as the spuriously oversized space of possible states (an increase in disorder) due to using the classification posterior in place of the indicator variable.
Whereas $H_{n}$ is minimized to a value of $0$ by any deterministic classification, $L_{n}$ is minimized to a value of $0$ only if $\tau_{n}$ and $p(m \mid d_{n})$ are equal to one another.
It can also be proven that the uncertain classifier of Section~\ref{sec:uncertaindata} maximizes $L_{n}$ \citep{murphy_machine_2012}.
As an aside, a difference between $L_{n}$ and $H_{n}$ evaluated at $\tau_{n, m}$ would be the information lost to disorder in using $p(m \mid d_{n})$ in place of $\tau_{n, m}$, also known as the Kullback-Leibler Divergence (KLD); see \citet{malz_approximating_2018} for a comprehensive exploration of the KLD for a continuous 1-dimensional probability space.

The log-loss has only recently established a presence in the astronomy literature \citep{hon_deep_2017, hon_deep_2018}.
Its greatest strength is that it is straightforwardly interpretable, enabling the metric itself to contribute to uncertainty propagation in an inference problem using the probability densities provided by the classifier.

\subsubsection{Brier score}
\label{sec:brier}

The Brier score \citep{brier_verification_1950}, given as
\begin{eqnarray}
  \label{eq:brier}
  B_{n}  \equiv Q^{B}_{n} &=& \sum_{m=1}^{M} (\tau_{n, m} - p(m \mid d_{n}))^{2},
\end{eqnarray}
is a mean square error calculated between the indicator variable and the classification posterior.
Unlike the log-loss, the Brier score has been used extensively in solar flare forecasting \citep{crown_validation_2012, mays_ensemble_2015, florios_forecasting_2018}, stellar variability identification \citep{richards_construction_2012, armstrong_k2_2016}, and star-galaxy separation \citep{kim_hybrid_2015}.

As with the log-loss, the Brier score is minimized to $0$ only for a perfect classifier.
The Brier score is an attractive option because it both rewards classifiers for  assigning more probability to the true class and penalizes classifiers for assigning any probability to classes other than the true class, in contrast to the log-loss, which only accounts for probability assigned to the true class.
We expect this difference to significantly distinguish the Brier score from the log-loss.

The interpretation of the Brier score is less obvious than that of the log-loss, as its dimensions depend on those of the probability space upon which the classification posteriors are defined.
In addition, modifying it with weights requires choosing whether to weight only per-object values $B_{n}$ or also the individual terms $B_{n, m}$ contributing to it.
We leave to future work the thorough investigation of a nontrivial weighting scheme on the Brier metric, however, opting to treat both metrics the same, according to the weighting scheme of Section~\ref{sec:weights}, in our implementation.

\subsection{Weights}
\label{sec:weights}

The most concerning systematics discussed in Section~\ref{sec:mockdata} are those of tunnel vision and cruise control.
The actual light curve data stream of \lsst\ will be particularly vulnerable to both due to extreme class imbalance and class hierarchy (for example different subtypes of a single transient or variable class).
This susceptibility is compounded by the nonrepresentativity of the \plasticc\ training set, which is designed to reflect the nonrepresentativity anticipated of \lsst.
Any metric under equal weight per light curve would incentivize tunnel vision and cruise control focused on the most prevalent class.
In order to meet the needs of science cases concerning other, rarer classes, \plasticc's metric will be more nuanced, even if it complicates the interpretability of the metric.

One option is to apply a threshold of classification efficacy on all classes in order to assign an overall winner, though it would require reducing the classification probabilities to deterministic class labels.
When doing binary classification with a method that reduces probabilities to deterministic class labels, each light curve is assigned the class of higher probability, even if the two probabilities are quite similar, a situation that is particularly likely if the light curve, in fact, belongs to a third class or if the two classes are subclasses of a single physical phenomenon.
A simple reduction to a deterministic label could be made more palatable with a secondary threshold mechanism.
For example, requiring a minimum difference in probability density between the maximum probability class and the next highest probability class would help avert this degeneracy.

A simpler alternative that we investigate in this paper is to use a weighted average
\begin{eqnarray}
  \label{eq:weightavg}
  Q &=& \frac{1}{\sum_{m} w_{m}} \sum_{m} w_{m} Q_{m}
\end{eqnarray}
of per-class metrics $Q_{m}$.
(While weights could be assigned to each term $Q_{n, m}$, we do not consider this complexity at this time.)
Weights that are not proportional to $N^{-1}$ nor $M^{-1}$ may be chosen to encourage challenge participants to direct more attention to classes with less active classification efforts or those that have been historically more difficult to classify due to observational limitations.

Downweighting the metrics of classes affected by counterproductive systematics could mitigate the impact of the tunnel vision or cruise control classifiers.
The weights for the \plasticc\ metric, however, must be determined before there is knowledge of which systematics affect which classes.
Because of this caveat, the choice of weights is isolated to an inherently human problem dictated by the value placed on the scientific merits of knowledge of each class.
This paper, on the other hand, can only quantify the impact of weights in relation to the systematics.
We thus agnostically test weighting schemes\footnote{The weights considered in this study are more extreme than those ultimately used for \plasticc\ because the true weights were blinded from some authors prior to the end of the challenge.
However, we note that the weights could be (and in fact were) discovered by \plasticc\ competitors by systematically probing the output of the public leader board with entries from the cruise control classifier archetype targeting each class one at a time.}
where classes affected by a particular systematic take a given weight $0 \leq w \leq 1$ and all other classes have a weight $(1 - w) / (M - 1)$.

\section{Results}
\label{sec:results}

In the following sections, we explore the response of the log-loss and Brier score metrics to the classifiers of Section~\ref{sec:data} and as a function of the weights on affected classes.

\subsection{Mock classifier systematics}
\label{sec:mockresults}

\begin{figure*}
	\begin{center}
		\includegraphics[width=0.99\textwidth]{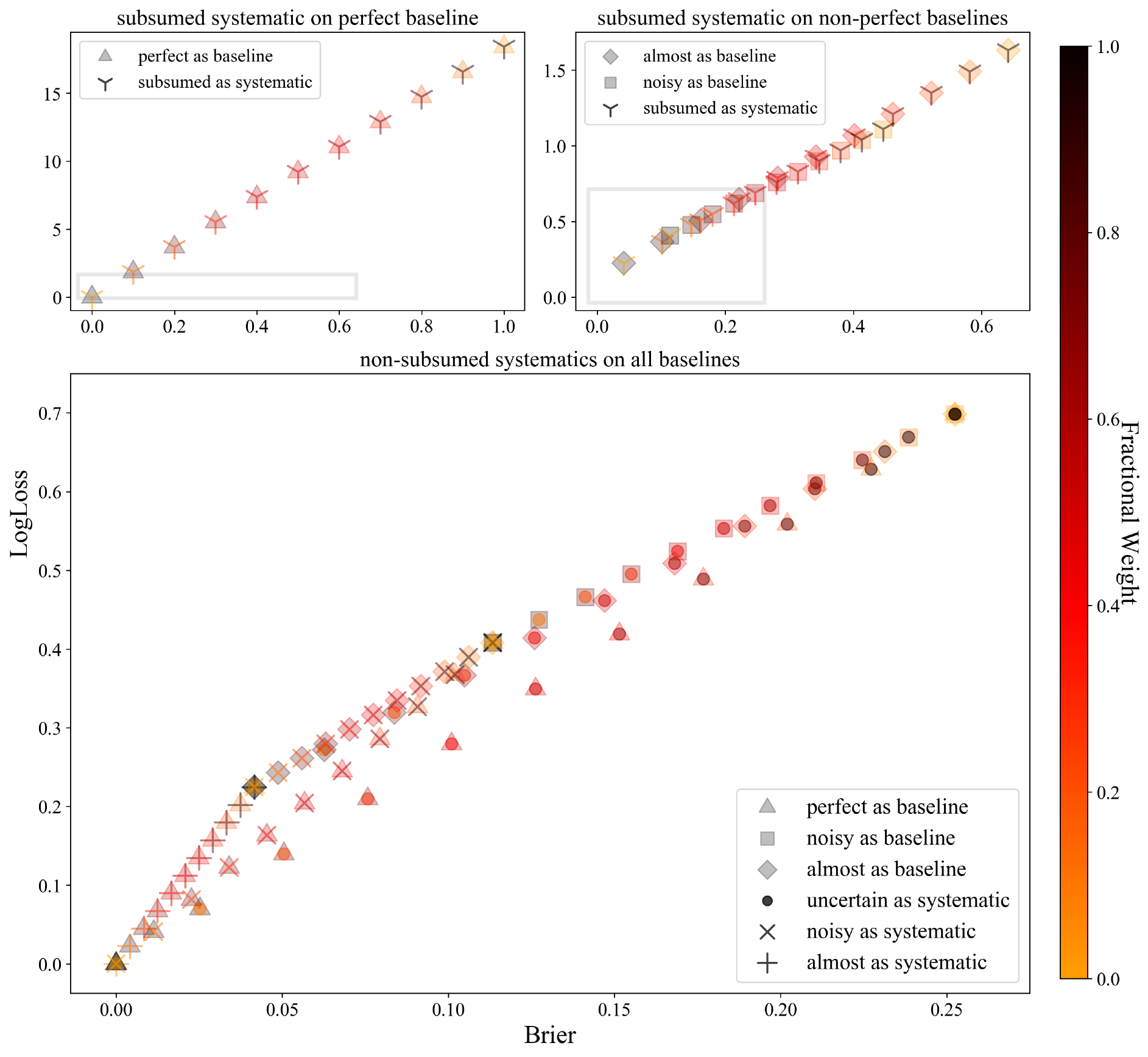}
		\caption{
		Weighted log-loss and Brier scores for baseline classifiers with combinations of systematics.
		Each point represents a classifier with a shared baseline behavior (regular polygon marker; triangle for perfect, diamond for almost perfect, square for noisy) for all but one class, which is affected by a particular systematic (asterisk markers; plus for almost perfect, cross for noisy, dot for uncertain, and Y-shape for subsumed).
		The color of the marker for the systematic effect indicates the weight on the one class affected by that systematic, while the color of the baseline behavior marker indicates the integrated weight evenly distributed over other classes with baseline behavior, where lower weights are greener and higher weights are bluer.
		From left to right, we zoom in on a particular range of scores, to highlight the scale of the effect of weighted systematics on the metrics for well-behaved methods with low Brier/log-loss values.
		The ranges of Brier score and log-loss values between the panels are in ratios of approximately 10:7:3 and 100:10:5, respectively, indicating the log-loss's higher sensitivity to the presence of systematics.
		The metrics are most sensitive to the subsuming systematic on a perfect baseline (triangle with Y-shaped marker), whereas other combinations of baseline and systematic can be grouped with a smaller dynamic range in both metrics.
		}
	\end{center}
	\label{fig:all_combined}
\end{figure*}

We simulate probabilistic classifications as potential submissions to \plasticc\ by the methodology of Section~\ref{sec:mockdata} based on CPMs composed of pairs of the characteristic classifiers shown in Figure~\ref{fig:all_combined} under various weightings described below.

The systematics introduced to each baseline are those that we intuitively expect to worsen classification performance of an arbitrary classifier:
\begin{itemize}
\item the uncertain, almost perfect, noisy, and subsuming classifiers are anticipated to worsen an otherwise perfect classifier;
\item the uncertain, noisy, and subsuming classifiers are anticipated to worsen an otherwise almost perfect classifier;
\item the uncertain and subsuming classifiers are anticipated to worsen an otherwise noisy classifier.
\end{itemize}
In every case, we apply the systematic to one true class, which corresponds to transforming one row of the baseline CPM.

The introduction of weights illustrates the effect each particular systematic has on a given baseline, and more importantly, how up- (or down-) weighting the affected class changes the overall metric value for the mock classifier.
Weighting schemes are defined by a weight $0 \leq w \leq 1$ on the affected class, with the remaining baseline classes sharing equal weight $(1 - w) / (M - 1)$; we test eleven weighting schemes with $w = 0., 0.1, \dots, 1.$.
A higher weight on the systematic corresponds to a lower weight on the more desirable baseline, causing both the log-loss and Brier score to increase.
This variation in weights establishes linear relationships between the log-loss and Brier score metrics for each pair of baseline and systematic, but the slope is related to the relative sensitivity of the metrics.

Figure~\ref{fig:all_combined} confirms that for all weight on the perfect classifier, the values of both metrics vanish to zero.
It is worth noting that the log-loss has more dynamic range than the Brier score overall, and that the log-loss is acutely sensitive to the subsuming systematic on a baseline of a perfect classifier.
However, the relative scales of metric values for different baseline-plus-systematic pairs are quite large, requiring three panels, zooming in from left to right.

The left panel of Figure~\ref{fig:all_combined} shows the largest variations in metric scores, for the combination of the perfect baseline and a subsuming systematic where one class is given a probability of 1 for being in another particular class and a probability of 0 for being in its true class.
This means both metrics are acutely sensitive to the subsuming systematic on a perfect baseline, which can only be overcome by aggressive downweighting.
In fact, the log-loss value for a classifier that subsumes a class into one that is classified perfectly should be infinite if the classes unaffected by the systematic have no weight; it is only finite for us because of the limits of numerical precision.

The middle panel of Figure~\ref{fig:all_combined} illustrates a narrower range of log-loss and Brier score for the subsuming systematic on the almost perfect and noisy classifier baselines.
The subsuming systematic on any baseline besides the perfect classifier defines a new regime of high but not infinite values of the metrics.

The right panel of Figure~\ref{fig:all_combined} shows the values for all other systematics on all baselines.
Though the slope is lower than in the other panels, the dynamic range of the log-loss remains higher; in other words, the log-loss is in general more sensitive to systematics than the Brier score.

In summary, both the log-loss and Brier score are most sensitive to the subsuming systematic than any other systematic.
Tuning the weights can provide an avenue toward imposing a global metric penalty on classifiers exhibiting a systematic on one class.

\begin{table}[]
\begin{tabular}{lll}
Classifier characteristic & Brier score & Log-loss\\
\hline
Perfect & 0.0 & 0.0\\
Almost perfect & 0.042 & 0.225\\
Noisy & 0.113 & 0.408\\
Uncertain & 0.253 & 0.699\\
Subsumed from Noisy & 0.447 & 1.109\\
Subsumed from Almost & 0.641 & 1.629\\
Subsumed from Perfect & 1.0 & 18.421\footnote{The entry for the log-loss of a classifier that subsumes a class into one that is otherwise perfectly classified should be infinite but is bounded by the numerical precision of our calculations.}
\end{tabular}
\caption{Metric values computed using Equation~\ref{eq:weightavg} with all weight on the mock class affected by the indicated systematic, described in Sec.~\ref{sec:mockdata}, corresponding to the $w=1$ cases in Figure~\ref{fig:all_combined}.
While the log-loss metric has a larger dynamic range than the Brier score for poor classification, the archetypical classifiers would be ranked (lower values are better) the same way by either metric.
}
\label{tab:extents}
\end{table}

When all weight is on the class exhibiting the systematic, there is a characteristic limit for each metric's values, shown in Table~\ref{tab:extents}.
Because a subsumed class takes the conditional probability vector of the subsuming class, the metric values depend on what systematics may be affecting the subsuming class as well.
While the two metrics obviously take different values, in accordance with their slopes given in Table~\ref{tab:slopes}, they do agree on the ranking of these classifiers.
Though this agreement is not in general guaranteed, it is a desirable behavior, indicating that these metrics would lead to the same conclusion about the severity of each systematic.

\begin{table}[]
\begin{tabular}{l|llll}
	& Systematics & & &\\
Baselines & Subsumed & Uncertain & Noisy & Almost\\
\hline
Perfect & 18.421 & 2.763 & 3.601 & 5.387\\
Almost perfect & 2.343 & 2.246 & 2.556 & \\
Noisy & 2.102 & 2.085 & &
\end{tabular}
\caption{
The slopes for each baseline-plus-systematic pair in the space of log-loss versus Brier score.
A higher slope corresponds to increased sensitivity of the log-loss over the Brier score to the systematic-baseline pair in question.
The contrast between log-loss and Brier score is highest on a baseline of the perfect classifier, meaning the log-loss may more strongly discriminate between classifiers that are already extremely good.
}
\label{tab:slopes}
\end{table}

The relative sensitivity ratios of the log-loss to the Brier score are the slopes in the trends of Figure~\ref{fig:all_combined} and are given in Table~\ref{tab:slopes}.
The log-loss always has higher sensitivity than the Brier score (i.e. it responds more strongly to up-weighting classes affected by a systematic), particularly to the difference between the perfect classifier and any lesser classifier.
A possible implication of this behavior is that the log-loss may have an enhanced ability to distinguish between multiple high-performing classifiers that might not have meaningfully different metric values under the Brier score.

On the other hand, the log-loss can be seen as more susceptible to the tunnel vision classifier because its value improves sharply with any move toward perfection.
If the subsumed class has little weight, the metric values are quite low, moreso for the log-loss than the Brier score.
This means that under a population-proportional weighting scheme, it would not be penalized for subsuming an uncommon class if it performed well for a more common class, a situation that would not serve the needs of the astronomical community.

\subsection{Representative classifications}
\label{sec:realresults}

\begin{figure}
	\begin{center}
		\includegraphics[width=0.49\textwidth]{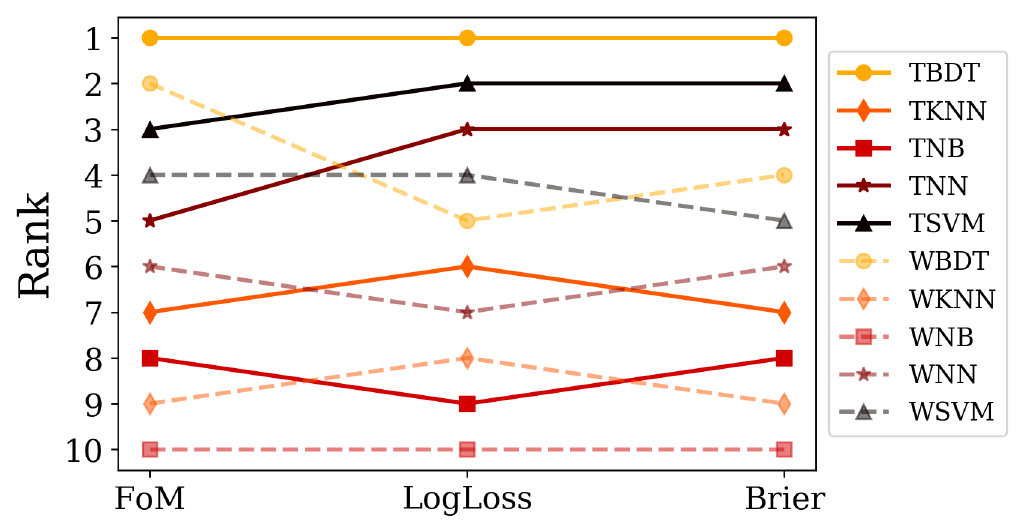}
		\caption{
		The rankings of each of the five \snmachine\ classification algorithms (Boosted Decision Tree (BDT), K-Nearest Neighbors (KNN), Naive Bayes (NB), Neural Network (NN), and Support Vector Machine (SVM)) on template (T*) and wavelet (W*) features with equal weight per object under the three metrics.
		The metrics broadly agree on the ranking of the classifiers, confirming consistency between a conventional metric of classification performance and the metrics of probabilistic classifications presented here.
		However, there are some differences with pairwise swapping between the log-loss and Brier rankings and some significant reordering of ranks 2 through 5 with the FoM metric relative to the probabilistic metrics.
		}
	\end{center}
	\label{fig:snmachineresults}
\end{figure}

We apply the log-loss and Brier metrics to the classification output from \snmachine.
While the classification methods described in \citet{lochner_photometric_2016} refer to the idealized subset of the \snphotcc\ data, these approaches are the state-of-the-art in classification of extragalactic transients.
We present in Figure~\ref{fig:snmachineresults} the rankings under the log-loss and Brier score metrics assuming an equal weight per object.

We apply our metrics to the classification output from \snmachine\ applied to the \snphotcc\ dataset as an example of representative light curves and representative classifiers used in extragalactic astronomy.
We present in Figure~\ref{fig:snmachineresults} the rankings of each classifier under the log-loss and Brier scores assuming an equal weight per object, as well as the original \snphotcc\ metric described in Section~\ref{sec:deterministic}.

The Brier score, log-loss, and \snphotcc\ FoM are in agreement as to the first- and last-ranked classifiers.
This consensus indicates that both of the potential \plasticc\ metrics are roughly consistent with our intuition about what makes a good classifier, providing an anchor between accepted notions of an appropriate metric and the metrics of probabilistic classifications under consideration here.
One should be careful not to generalize, however, as the rankings under the three metrics are not identical.

We note that the FoM differs more from the Brier score and log-loss metrics than they do from one another.
This is perhaps unsurprising, given that the \snphotcc\ was specifically looking to value classification algorithms that were pure (that yielded a large number of SNIa classifications and few interlopers from the other classes), as opposed to metric that rewards good performance across classes.

\section{Discussion}
\label{sec:discussion}

The goal of this work is to identify the metric most suited to \plasticc, which seeks classification posteriors of complete light curves similar to those anticipated from \lsst, with an emphasis on classification over all types, rewarding a ``best in show'' classifier rather than focusing on any one class or scientific application.\footnote{At the conclusion of \plasticc, other metrics specific to scientific uses of one or more particular classes will be used to identify ``best in class'' classification procedures that will be useful for more targeted science cases.}
The weighted log-loss is thus the metric most suited to the current \plasticc\ release.

Because transient and variable object classification is crucial for a variety of scientific objectives, the impact of a shared performance metric on this diversity of goals leads to complex and covariant trade-offs.
Though the selection criteria for metrics specific to each science goal are outside the scope of this work, which concerns only the first instantiation of \plasticc, we discuss below some issues concerning the identification of metrics for a few example science cases.

\subsection{Ongoing transient follow-up}
\label{sec:early}

Spectroscopic follow-up is only expected of a small fraction of \lsst's detected transients and variable objects due to limited resources for such observations.
In addition to optical spectroscopic follow-up, photometric observations in other wavelength bands (near infrared and x-ray from space; microwave and radio from the ground) or at different times will be key to building a physical understanding of the object, particularly as we enter the era of multi-messenger astronomy with the added possibility of optical gravitational wave signatures.
Prompt follow-up observations are highly informative for fitting models to the light curves of familiar source classes and to characterizing anomalous light curves that could indicate never-before-seen classes that have eluded identification due to rarity or faintness.
As such, decisions about follow-up resource allocation must be made quickly and under the constraint that resources wasted on a misclassification consume the budget remaining for future follow-up attempts.
A future version of \plasticc\ focused on early light curve classification should have a metric that accounts for these limitations and rewards classifiers that perform better even when fewer observations of the lightcurve are available.

We consider the decision of whether to initiate follow-up observations to be binary and deterministic.
However, it is possible to conceive of non-binary decisions about follow-up resources; for example, one could choose between dedicating several hours on a spectroscopic instrument following up on one likely candidate or dedicating an hour each on several less likely candidates.
Here, we will discuss a metric for an early classification challenge to be focused on deterministic classification because the conversion between classification posteriors and decisions is uncharted territory that we do not explore at this time.

Even within the scope of spectroscopic follow-up as a primary motivation for early light curve classification, the goals of model-fitting to known classes and discovery of new classes would likely not share an optimal metric.
The critical question for choosing the most appropriate metric for any specific science goal motivating follow-up observations is to maximize information.
We provide two examples of the kind of information one must maximize via early light curve classification and the qualities of a deterministic metric that might enable it.

\subsection{Spectroscopic supernova cosmology}
\label{sec:spec_sncosmo}

Supernova cosmology with spectroscopically confirmed light curves benefits from true positives, which contribute to the constraining power of the analysis by including one more data point;
when the class in which one is interested is as plentiful as SN Ia and our resources limited a priori, we may not be concerned by a high rate of false negatives.
False positives, on the other hand, may not enter the cosmology analysis, but they consume follow-up resources, thereby depriving the endeavor of the constraining power due to a single SN Ia.

A perfect classifier would lead to a maximum amount of information about the cosmological parameters conditioned on the follow-up resource budget.
Consider deterministic labels derived from cutoffs in probabilistic classifications for this scientific application; raising the probability cutoff reduces the number of false positives, boosting the cosmological constraining power, but at a cost of increasing the number of false negatives, which represent constraining power forgone.
As this tradeoff is asymmetric, it is insufficient to consider only the true and false positive and negative rates, as the \snphotcc\ FoM does, without propagating their impact on the information gained about the cosmological parameters.

\subsection{Anomalous transient and variable detection}
\label{sec:anom}

A particularly exciting science case is anomaly detection, the discovery of entirely unknown classes of transient or variable astrophysical sources, or distinguishing some of the rarest types of sources from more abundant types.
Like the case of spectroscopic supernova cosmology discussed above, anomaly detection also gains information only from true positives, but the cost function is different in that the potential information gain is unbounded when there is no prior information about undiscovered classes.
The discovery of pulsars serves as an example of novelty detection enabled by a human classifier \citep{hewish_observation_1968, bell_burnell_measurement_1969}.

Resource availability for identifying new classes is more flexible, increasing when new predictions or promising preliminary observations attract attention, and decreasing when a discovery is confirmed and the new class is established.
In this way, a false positive does not necessarily consume a resource that could otherwise be dedicated to a true positive, and the potential information gain is sufficiently great that additional resources would likely be allocated to observe the potential object.
Thus, a metric for evaluating anomaly detection effectiveness would aim to minimize the false negative rate and maximize the true positive rate.

\subsection{Difficult light curve classification}
\label{sec:difficult}

Photometric light curve classification may be challenging for a number of reasons, including the sparsity and irregularity of observations, the possible classes and how often they occur, and the distances and brightnesses of the sources of the light curves.
These factors may represent limitations on the information content of the light curves, but appropriate classifiers may be able to overcome them to a certain degree.

Though quality cuts can eliminate the most difficult light curves from entering samples used for science applications, such a practice discards information that may be of value under an analysis methodology leveraging the larger number of light curves included in a sample without cuts.
Thus, classification methods that perform well on light curves characterized by lower signal-to-noise ratios are specially important for exploiting the full potential of upcoming surveys like \lsst.

This version of \plasticc\ implements quality cuts to homogenize difficulty to some degree, and notions of classification difficulty may depend on information that will not be available until after the challenge concludes.
While the groundwork for a metric incorporating data quality has been laid by \citet{wu_radio_2018}, we defer to future work an investigation of this possibility.

\section{Conclusion}
\label{sec:conclusion}

As part of the preparation for \plasticc\, we investigated the properties of metrics suitable for probabilistic light curve classifications in the absence of a single scientific goal.
Therefore, we sought a metric that avoids reducing classification probabilities to deterministic labels and is compatible with a multi-class, rather than binary (two-class), setting.
We did not consider some of the most popular metrics used in astronomy (such as accuracy, combinations of the true and false positive and negative rates, and AUC functions thereof) because they did not satisfy these criteria, even though it is in principle possible to extend such metrics for our situation.
Our experimental design thus explores the response of potential metrics to simulated classification submissions from a set of mock classifier archetypes expected of generic transient and variable classifiers.

We identified two metrics of multi-class classification probabilities established in the literature: the Brier score and the log-loss.
The Brier score and the log-loss metrics are structurally and conceptually different, with wholly different interpretations.
The Brier score is a sum of square differences between probabilities;
the explicit penalty term is an attractive feature, but it treats probabilities as generic scores.
The log-loss on the other hand is readily interpretable as a measure of information, meaning the metric itself could be propagated into forecasting the cosmological constraining power of \lsst, affecting the choice of observing strategy.

When evaluated with equal weight on each classified object, both the Brier score and the log-loss metrics are susceptible to rewarding a classifier that performs well on the most prevalent class and poorly on all others, which fails to meet the needs of \plasticc's diverse motivations under the unavoidable population imbalances of astronomical data.
To discourage competitors from neglecting rare classes, we explored a weighted average of the metric values on a per-class basis as a possible mitigation strategy to incentivize classifying uncommon classes, effectively ``leveling the playing field'' in the presence of highly nonuniform class membership.

On the basis of the mock classifier rankings, we found that both metrics reward the classifiers that are better and penalize those that are worse, where better and worse are defined by our common intuition, yielding the same rankings under either metric and demonstrating that both could be appropriate for \plasticc.
However, since only one could be selected, the log-loss was chosen due to its potential for interpretation after the conclusion of the challenge.
While modifyinging the log-loss metric to handle weights for different classes diminishes its interpretability, it can still be understood as information gain, subject to the value we as scientists place on knowledge of each class.

The space of possible metrics we could have considered is truly unbounded, from traditional metrics of deterministic labels to established extensions thereof for probabilistic classifications to novel quantities tuned to any given science case.
Though there was no need to do a more extensive survey of metrics nor to devise new metrics for \plasticc, since both log-loss and Brier score passed the basic sanity tests for this application, further work remains to be done in optimally selecting probabilistic classification metrics in other astronomical contexts.

We conclude by noting that care should be taken in planning future open challenges to ensure alignment between the challenge goals and the performance metric, so that efforts are best directed to achieve the challenge objectives.
It is our hope hope that this study of metric performance across a range of systematic effects and weights may serve as a guide to approaching the problem of identifying promising probabilistic classifiers for general science applications.


\subsection*{Acknowledgments}

Author contributions are listed below. \\
A.I.~Malz: conceptualization, data curation, formal analysis, investigation, methodology, project administration, software, supervision, validation, visualization, writing - editing, writing - original draft \\
R.~Hlo\v{z}ek: data curation, formal analysis, funding acquisition, investigation, project administration, software, supervision, validation, visualization, writing - editing, writing - original draft \\
T.~Allam Jr: investigation, software, validation, writing - original draft \\
A.~Bahmanyar: formal analysis, investigation, methodology, software, writing - editing, writing - original draft \\
R.~Biswas: conceptualization, methodology, software, supervision, writing - editing, writing - original draft \\
M.~Dai: writing - editing \\
L.~Galbany: writing - editing \\
E.E.O.~Ishida: conceptualization, project administration, supervision, writing - editing \\
S.W.~Jha: writing - editing \\
D.~Jones: software \\
R.~Kessler: writing - editing \\
M.~Lochner: conceptualizaton, data curation, formal analysis, visualization, writing - editing \\
A.A.~Mahabal: data curation, software, writing - editing, writing - original draft \\
K.S.~Mandel: conceptualization, supervision, writing - editing \\
J.R.~Mart\'inez-Galarza: data curation, software, visualization, writing - original draft \\
J.D.~McEwen: conceptualization, investigation, supervision \\
D.~Muthukrishna: data curation, validation \\
G.~Narayan: data curation, formal analysis \\
H.~Peiris: conceptualization, funding acquisition, supervision \\
C.M.~Peters: writing - editing \\
K.~Ponder: visualization, writing - editing \\
C.N.~Setzer: conceptualization, software \\

This paper has undergone internal review in the LSST Dark Energy Science Collaboration. 
The authors would like to thank Melissa Graham, Weikang Lin, and Chad Schafer for serving as the LSST-DESC publication review committee, as well as Tom Loredo for other helpful feedback.
The authors also express gratitude to the anonymous referee for substantive suggestions that improved the paper.

\software{
jupyter \citep{kluyver_jupyter_2016},
matplotlib \citep{hunter_matplotlib:_2007},
numpy \citep{oliphant_guide_2006, oliphant_python_2007, walt_numpy_2011},
proclam \citep{malz_proclam_2018},
scikit-learn \citep{pedregosa_scikit-learn:_2011},
scipy \citep{jones_scipy:_2001, buitinck_api_2013}
}

AIM was advised by David W. Hogg and was supported by National Science Foundation grant AST-1517237.
TA is supported in part by STFC.
RB and CS are supported by the Swedish Research Council (VR) through the Oskar Klein Centre.
Their work was further supported by the research environment grant ``Gravitational Radiation and Electromagnetic Astrophysical Transients (GREAT)'' funded by the Swedish Research council (VR) under Dnr 2016-06012.
AAM was supported in part by the NSF grants AST-0909182, AST-1313422, AST-1413600, and AST-1518308, and by the Ajax Foundation

The financial assistance of the National Research Foundation (NRF) towards this research is hereby acknowledged.
Opinions expressed and conclusions arrived at, are those of the authors and are not necessarily to be attributed to the NRF.
This work is partially supported by the European Research Council under the European Community’s Seventh Framework Programme (FP7/2007-2013)/ERC grant agreement no 306478-CosmicDawn.

Canadian co-authors acknowledge support from the Natural Sciences and Engineering Research Council of Canada.
The Dunlap Institute is funded through an endowment established by the David Dunlap family and the University of Toronto.
The authors at the University of Toronto acknowledge that the land on which the University of Toronto is built is the traditional territory of the Haudenosaunee, and most recently, the territory of the Mississaugas of the New Credit First Nation.
They are grateful to have the opportunity to work in the community, on this territory.

We acknowledge the University of Chicago Research Computing Center for support of this work.
This research used resources of the National Energy Research Scientific Computing Center (NERSC), a U.S. Department of Energy Office of Science User Facility operated under Contract No. DE-AC02-05CH11231.
This research at Rutgers University is supported by US Department of Energy award DE-SC0011636.

The DESC acknowledges ongoing support from the Institut National de Physique Nucl\'eaire et de Physique des Particules in France; the Science \& Technology Facilities Council in the United Kingdom; and the Department of Energy, the National Science Foundation, and the LSST Corporation in the United States.  DESC uses resources of the IN2P3 Computing Center (CC-IN2P3--Lyon/Villeurbanne - France) funded by the Centre National de la Recherche Scientifique; the National Energy Research Scientific Computing Center, a DOE Office of Science User Facility supported by the Office of Science of the U.S.\ Department of Energy under Contract No.\ DE-AC02-05CH11231; STFC DiRAC HPC Facilities, funded by UK BIS National E-infrastructure capital grants; and the UK particle physics grid, supported by the GridPP Collaboration.  This work was performed in part under DOE Contract DE-AC02-76SF00515.

\bibliography{main,lsstdesc}

\end{document}